# Development of a thin-target hard X-ray bremsstrahlung detection system to study confined runaway electrons in Aditya-U Tokamak

Suman Dolui[1,2] , Santosh Pandya[1], J.Kumar[1], Bharat Hegde[1,2], Kaushlender Singh[3], J.Ghosh[1,2], Komal Yadav[1,2], Mitul Abhangi[1], Shishir Purohit[1], Minsha Shah[1], Laxmikanta Pradhan[4], Harshita Raj[1,2], R.L. Tanna[1], Ashok K. Kumawat[1,2], Injamul Hoque[1,2], Soumitra Banerjee[1,2], Ruchi Varshney[1,2] , S.Aich[1], Rohit Kumar[1], K. A. Jadeja[1] , K. M. Patel[1], M.K. Gupta[1], P. K. Chattopadhyay[1,2], A.Sen[1,2], Y.C. Saxena[1,2], R.Pal[5]

[1] *Institute for Plasma Research, Gandhinagar 382 428*
[2] *Homi Bhabha National Institute (HBNI), Mumbai 400 085*
[3] *EPFL, Route Cantonale, 1015 Lausanne, Switzerland*
[4] *Indian Space Research Organisation, Bengaluru 560094*
[5] *Saha Institute for Nuclear Physics, Kolkata 700 064*

A specially shielded CdTe detector based hard X-ray (HXR) monitoring system equipped with a lead collimator has been developed and installed on the Aditya-U tokamak to investigate the dynamics of fast electrons (~20–200 keV) generated during sawtooth activity. The pre-existing HXR monitor in Aditya-U is exposed to the entire HXR bremsstrahlung emission from the plasma volume, peripheral limiters, and other structural components, which limits its ability to separately study the dynamics of lost and confined runaway electrons (REs). In contrast, the newly developed diagnostic has successfully measured the chord-averaged thin-target HXR bremsstrahlung emission encompassing the core plasma region, particularly within and around the sawtooth inversion radius. The measured HXR spectra are validated through forward modeling code that incorporates plasma parameters, confined RE characteristics, and the geometric configuration of the diagnostic system. The results confirm the capability of the developed HXR monitor to probe the fast-electron dynamics during internal plasma instabilities.

## I. INTRODUCTION:

The bursts of HXR spikes during the sawtooth crashes[1] is a very commonly observed phenomena in Tokamaks[2–4] and the Aditya-U Tokamak is no exception[5] . Experimental observations in several tokamaks indicate that sawtooth crashes may be associated with nonthermal electron beams (15–150 keV) generated during magnetic reconnection near the X-points of magnetic islands. In DIII-D, the HXR bursts are observed to shift toward larger minor radii across the plasma column during the crash[6]. However, the lack of internal magnetic field measurements prevents a direct link between the formation of nonthermal beams and magnetic reconnection. Moreover, since sawtooth crashes do not always occur through magnetic reconnection [78910], this challenges the conventional view that nonthermal electrons are generated solely by the strong parallel electric fields produced during magnetic reconnection. These observations suggest that two concurrent processes may occur during sawtooth crashes: the loss of pre-existing runaway beams [111213] and the generation of new beams during reconnection events. This ambiguity cannot be resolved through observations of HXR bursts during a sawtooth crash using a tangential HXR monitor, which contain contributions from both thin-target bremsstrahlung [1415] emitted by energetic electrons interacting with the bulk plasma and thick-target bremsstrahlung produced when these electrons are lost to surrounding limiters and support structures. To overcome this limitation, a new HXR monitor has been developed with a line-of-sight restricted to the core plasma region, including the sawtooth inversion radius[10], while minimizing sensitivity to emissions from plasma-facing components. This geometry effectively suppresses the detection of thick-target bremsstrahlung arising from limiter interactions. Simultaneous measurements with the newly developed core-viewing HXR monitor and the existing tangential HXR monitor will enable discrimination between core-generated thin-target emission and limiter-associated thick-target emission. Such synchronized observations are expected to address the long-standing question in Aditya-U regarding whether the sawtooth crash leads to the generation of a non-thermal electron beam in the plasma core or instead triggers the loss of a pre-existing non-thermal electron population to the limiter.

The line-of-sight restricted HXR diagnostic[16,17]measures thin-target bremsstrahlung emission emitted radially across the ambient toroidal magnetic field. The diagnostic assembly is mounted above the top port at the vessel centerline, enabling a vertically downward view of the plasma at a selected poloidal plane. Section II presents a detailed description of the diagnostic setup. The experimentally measured HXR spectra are further verified using a forward-modelling code. Section III provides a comparison between the experimental results and the spectra obtained from the forward-modelling calculations. Finally, Section IV presents the conclusions along with the future objectives of this work.

## II. EXPERIMENTAL ARRANGEMENT:

The experiments were carried out in the Aditya-U Tokamak[18], which has a major radius $R = 0.75\ m$ and a minor radius $= 0.25m$ . The machine can operate with a maximum toroidal magnetic field of $1.4\ T$ and a maximum plasma current of $250\ kA$. A maximum plasma discharge duration of approximately $400\ ms$ has been achieved. The typical plasma density and electron temperature of the discharges are in the ranges of $1 - 3 \times 10^{19} m^{-3}$ and $300 - 500\ eV$, respectively. The limiter configuration consists of two crescent-shaped poloidal limiters installed on the outer (low-field) side of the vessel and a continuous toroidal belt limiter located on the inner (high-field) side of the vacuum vessel, which defines the plasma minor radius. The HXR detector used as the primary diagnostic in the present experiment consists of a Sodium Iodide (Tl) scintillation detector ($76\ mm$ diameter and $76\ mm$ thickness) coupled to a photomultiplier tube [13]. The photomultiplier output signal is fed into a spectroscopic amplifier with a shaping time constant of $0.5\ \mu s$, resulting in a total pulse width of approximately $4\ \mu s$ for a single photon event. The amplifier output is subsequently digitized directly for further analysis. The detector is positioned approximately $10\ m$ away from the center of the torus at a vertical height corresponding to the mid-plane of the machine. With this viewing geometry, the detector line-of-sight encompasses the entire plasma volume as well as the peripheral poloidal and toroidal limiters[5].

### a. Thin-target hard X-ray bremsstrahlung detection diagnostic:

The thin-target bremsstrahlung detection system consists of a CdTe-based detector[19]. integrated with a collimator, lead shielding, and appropriate filtering components, along with associated electronics such as a preamplifier, pulse shaper, and NIM bin module. To ensure a restricted field of view and, the entire detection assembly is positioned in close proximity to the source, plasma. Operation in a strong magnetic environment imposes significant constraints on detector selection. Conventional scintillation detectors, typically coupled with photomultiplier tubes (PMTs), are highly sensitive to magnetic fields. This sensitivity necessitates the use of bulky magnetic shielding, which complicates system design and limits compact arrangement. To address this limitation, a semiconductor-based CdTe detector [20,21] has been employed for hard X-ray (HXR) detection. CdTe detectors offer several advantages, including compact dimensions ($5 \times 5 \times 2\,mm^3$) and intrinsic insensitivity to magnetic fields, making them well suited for operation in magnetically harsh environments. In addition to their robustness, CdTe detectors exhibit several favorable characteristics for HXR measurements. These include room-temperature operation, high count-rate capability (up to several hundred kHz), and good energy resolution (typically $6 - 8\,keV$ at $122\,keV$). Furthermore, the high atomic numbers of cadmium ($Z = 48$) and tellurium ($Z = 52$) contribute to strong X-ray absorption efficiency. The X-ray absorption performance of CdTe crystals of varying thicknesses has been evaluated using mass attenuation coefficients obtained from the NIST XCOM database. The results, presented in Fig. 5, indicate that CdTe detectors provide moderate detection efficiency in the photon energy range of $30 - 300\,keV$, which is particularly suitable for studying thin-target bremsstrahlung produced by non-thermal electron beams. A comparative analysis of crystal thicknesses ($1\,mm, 2\,mm, and\ 3\,mm$) shows that increasing thickness enhances absorption efficiency at higher photon energies. However, thicker crystals also introduce higher noise levels, increased dead time, and require elevated bias voltages. Based on these trade-offs, a thickness of $2\,mm$ has been selected as the optimal configuration. The detector is enclosed within a $0.5\,mm$ thick aluminum housing, which serves to attenuate low-energy X-rays (below $10\,keV$) and reduce background noise. Additionally, a $1\,mm$ thick aluminum foil is used as an external window to ensure minimal distortion of the incident X-ray energy spectrum. Consequently, a total effective aluminum filter thickness of $1.5\,mm$ is implemented to suppress low-energy contributions while maintaining the integrity of the measured hard X-ray spectrum. A key objective of the present experiment is the measurement of thin-target bremsstrahlung originating from the plasma core. To ensure the fidelity of these measurements, particular attention has been given to minimizing contributions from hard X-rays generated at the limiter or arising due to scattering from the vessel walls. To achieve this, the detector assembly is enclosed within a cylindrical lead shield of $6\,cm$ thickness and $32\,cm$ length. This configuration is specifically designed to attenuate thick-target bremsstrahlung radiation. As illustrated in Fig. 1a, the transmission characteristics of lead for varying thicknesses indicate that a $6\,cm$ thick shield provides substantial attenuation for photon energies up to approximately $500\,keV$. Although higher-energy photons may still penetrate the shielding, their contribution at the detector location is expected to be minimal. This is because high-energy bremsstrahlung photons are emitted anisotropically, with peak intensity aligned along the electron beam direction[22], resulting in a reduced flux of photons above $500\,keV$ at the detector position. Consequently, the selected shielding thickness is considered sufficient for suppressing high-energy background radiation. Additionally, the shielding effectively reduces secondary signals arising from scattered photon flux. The detector is fully enclosed within the lead shielding, except for a collimated line-of-sight defined by a lead collimator positioned at the front as illustrated in Fig.1b. The collimator has an outer diameter of $85\,mm$, an inner aperture of $8\,mm$ (pinhole), and a length of $32\,cm$. Its dimensions are matched to the inner diameter of the shielding pipe to ensure a compact and well-aligned assembly. The detector is positioned $2\,cm$ behind the exit aperture of the collimator to optimize photon acceptance. The complete detector and shielding assembly is mounted on the top port of the vacuum vessel, aligned along the central vertical axis. The geometry of the collimator specifically, its $8\,mm$ aperture and $32\,cm$ length defines a narrow field of view directed vertically downward along the vessel central line[2324]. The collimator is placed in direct contact with the fused silica window of the top flange, ensuring that incident γ-photons enter directly through the window and pass into the collimator. This configuration minimizes the contribution of secondary photons generated by interactions with the stainless steel vessel walls, thereby improving the fidelity of the measured signal. A corresponding 63 CF flange with a glass window of approximately $7\ cm$ diameter is located at the bottom port, directly opposite (180°) the top port. The collimator geometry has been carefully designed such that its field of view is restricted entirely to this bottom window, thereby eliminating exposure to vessel walls. This alignment was experimentally verified by placing a white sheet above the bottom port inside the vessel and illuminating it from the detector side. The observation of a well-defined circular illumination pattern of approximately $7\ cm$ diameter confirmed the accuracy of the collimator design.

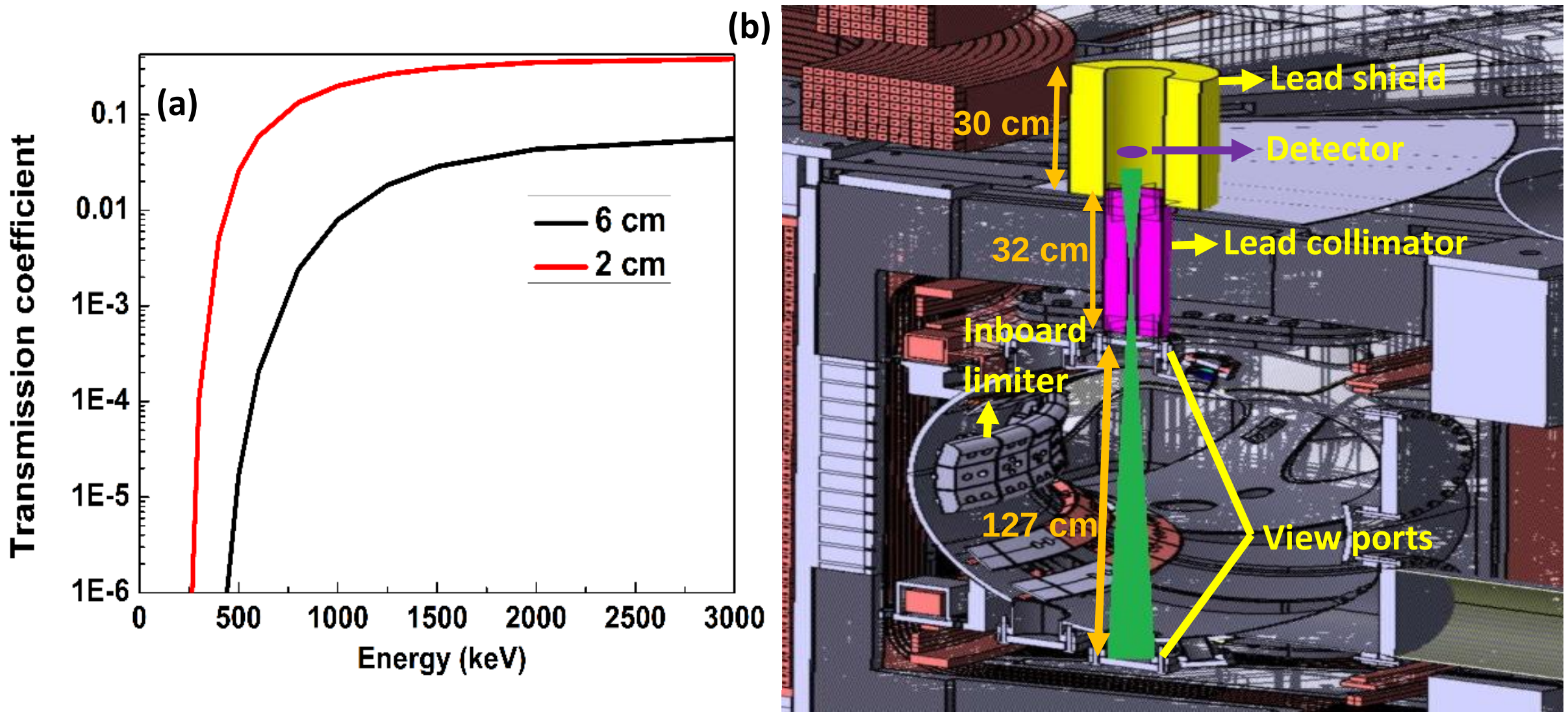

**Figure 1.** (a) Transmission coefficient of lead as a function of material thickness. (b) Schematic diagram of the experimental setup used for thin-target bremsstrahlung measurement.

### b. Suitable electronics required for the diagnostic

The energy deposited by incident hard X-ray (HXR) photons in the CdTe detector is converted into an electrical signal, which is subsequently processed and recorded using a data acquisition system. To minimize electromagnetic interference, the associated front-end electronics comprising a charge-sensitive preamplifier, pulse shaper, and amplifier are positioned at a distance from the detector assembly. A schematic representation of the primary electronic configuration is illustrated in Fig. 2 .The detector is typically operated with a bias voltage in the range of $100 - 150\ V$. Upon interaction of X-rays with the CdTe crystal, electron–hole pairs are generated, producing charge signals that are converted into voltage pulses by a charge-sensitive preamplifier. The preamplifier, with a sensitivity of $0.24\ mV/keV$, is well matched to the detector's photon energy response. It features a sufficiently short decay (fall) time of $50\ \mu s$ (PR-16 model), enabling operation at high count rates, while maintaining a low intrinsic noise level of approximately $3\ keV$. The maximum count rate capability of the system is $\sim 3 \times 10^5$ counts per second (cps), primarily limited by the charge collection time within the detector volume. At high count rates, particularly for high-energy photon fluxes, the preamplifier output may exhibit saturation effects accompanied by a DC baseline shift. The preamplifier output is therefore further processed using a pulse shaping amplifier (CEAN), which conditions the signal for accurate analysis. The shaper output is configured to produce Gaussian-shaped pulses, with adjustable shaping times ranging from $0.5 - 6\ \mu s$. For high count-rate applications, shorter shaping times ($0.5 - 1\ \mu s$) are preferred to optimize pulse height analysis (PHA) while minimizing pulse pile-up. The shaped signals are digitized using a fast analog-to-digital converter (ADC) operating at $1\ MHz$ and subsequently recorded for offline analysis. Post-processing of the acquired data enables flexible binning in both time and energy domains. A custom Python-based algorithm has been developed for peak detection and energy binning, facilitating efficient pulse height analysis. The use of high-speed digitization further reduces system dead time, thereby enhancing the overall counting performance. Energy calibration of the detector is carried out using a $Ba^{133}$ radioactive source with an activity of approximately $1\ \mu Ci$. This source is selected due to its multiple gamma emission lines spanning the energy range of $30 - 360\ keV$, enabling calibration across the detector's operational spectrum using a single reference. During calibration, the detector is biased at $130\ V$, and the amplifier gain is set to $200$. Signal acquisition is performed using an oscilloscope with a sampling rate of $5\ MHz$, and data collected over a $20\ s$ interval is used for pulse height analysis. The recorded count statistics are sufficient to generate well-defined energy spectra and histograms. The primary objective of the calibration procedure is to establish a reliable relationship between the measured pulse amplitude (voltage) and the corresponding incident photon energy, thereby enabling accurate spectral characterization of the detected HXR radiation.

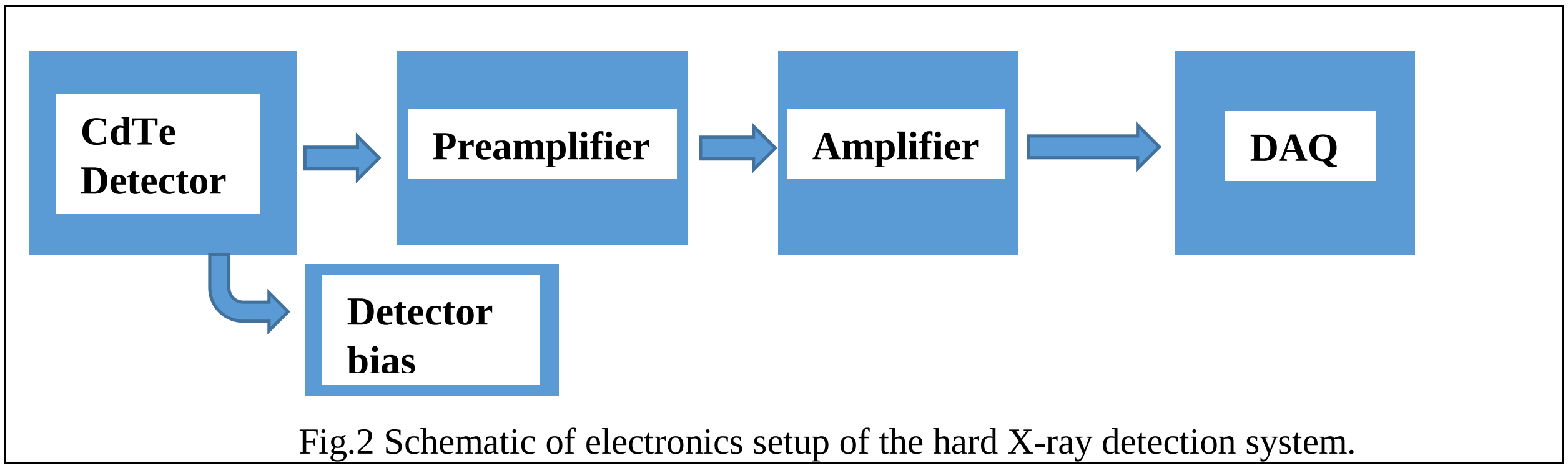

Fig.2 Schematic of electronics setup of the hard X-ray detection system.

## III. EXPERIMENTAL RESULTS:

Figure 3 illustrates a representative Ohmic discharge (#38965) in the Aditya-U tokamak. During the flat-top phase, the loop voltage is maintained in the range of $\sim 1 - 2\ V$, with a peak plasma current of 133 kA and a maximum central chord-averaged electron density of $1.32 \times 10^{19} m^{-3}$. The hard X-ray (HXR) signal shown in the figure is measured using a NaI (Tl) scintillation detector[16], which measures both thin-target and thick-target bremsstrahlung emissions. The detector operates in counting mode to record incident photon events. However, under high photon flux conditions, the event rate exceeds the processing capability of the associated electronics. As a result, pulse pile-up becomes significant, leading to distortion of the signal. This effect manifests as a progressive shift of the output toward negative values, ultimately causing signal saturation. Such behavior indicates the limitation of the detection system in handling high count rates, thereby restricting its effectiveness for accurate HXR measurements under these operating conditions.

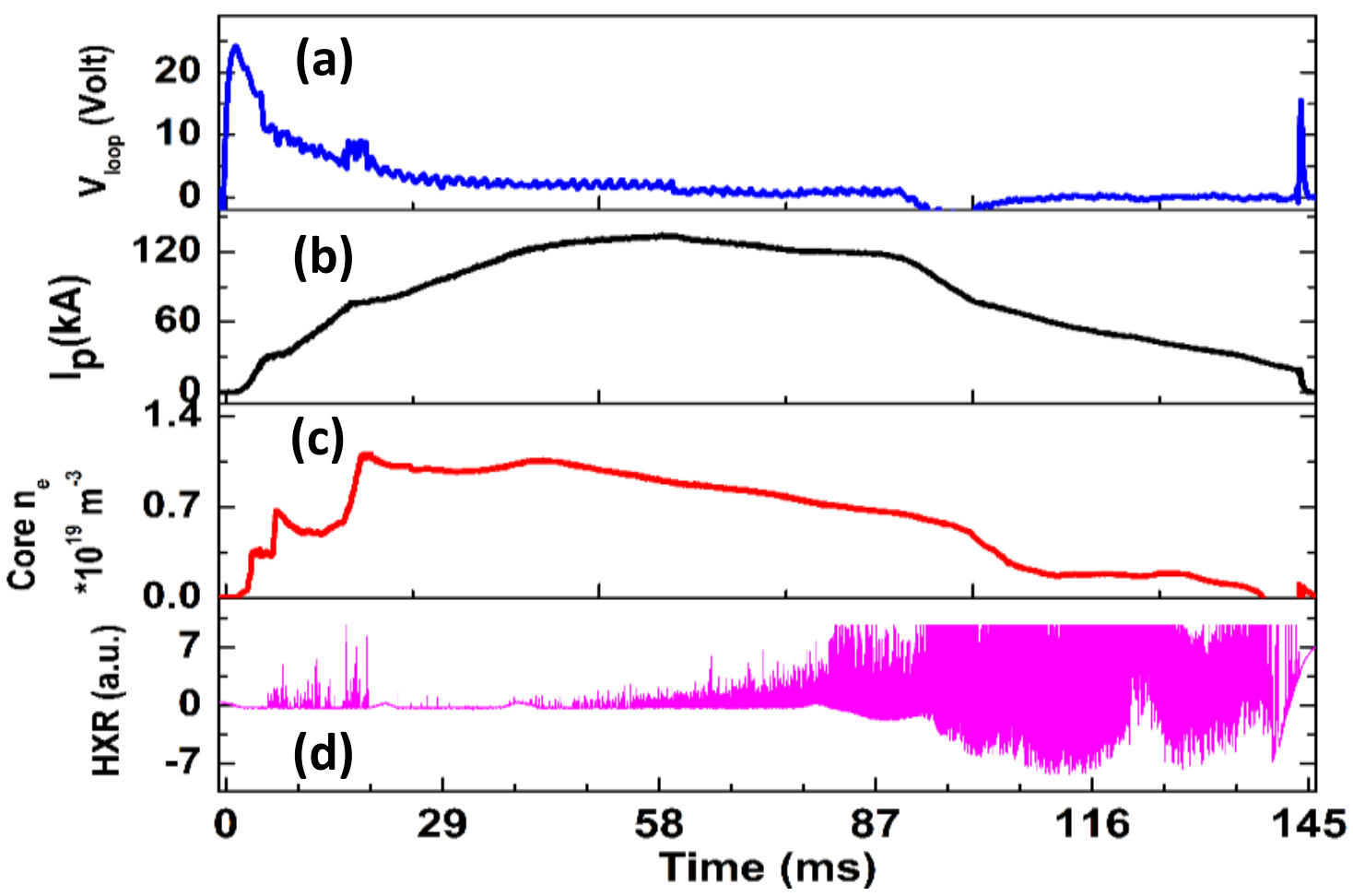

**Fig. 3.** Typical discharge parameters of Aditya-U: (a) loop voltage, (b) plasma current, (c) line-averaged central chord electron density, and (d) hard X-ray (HXR) emissivity including contributions from both thick- and thin-target bremsstrahlung emission measured by NaI(Tl) detector.

The hard X-ray (HXR) signal measured using the CdTe detector through the lead collimator is presented in Fig. 4, acquired in spectroscopic mode. To verify that the detected signal originates exclusively from radiation entering through the collimator aperture, a comparative analysis was performed using two plasma discharges, discharge numbered 38965 and 38986, which exhibit closely matched plasma parameters. The anode current signals from the photomultiplier tube (PMT) coupled to the NaI scintillation detector are comparable in both discharges prior to saturation, confirming that the event rates are effectively equivalent (see Fig. 4b). In discharge 38986, the collimator aperture was fully obstructed using a lead rod, whereas in discharge 38965, the aperture remained open. Under the blocked condition, the HXR signal is observed to nearly vanish, as indicated by the red trace in Fig. 4c. In contrast, a clear and significant signal is detected when the aperture is open, as shown by the black trace. This marked difference demonstrates that the measured HXR signal in the open configuration arises predominantly from bremsstrahlung emission transmitted through the collimator aperture. These results provide strong experimental validation that the newly developed HXR diagnostic system effectively isolates and detects bremsstrahlung radiation originating from the targeted plasma volume.

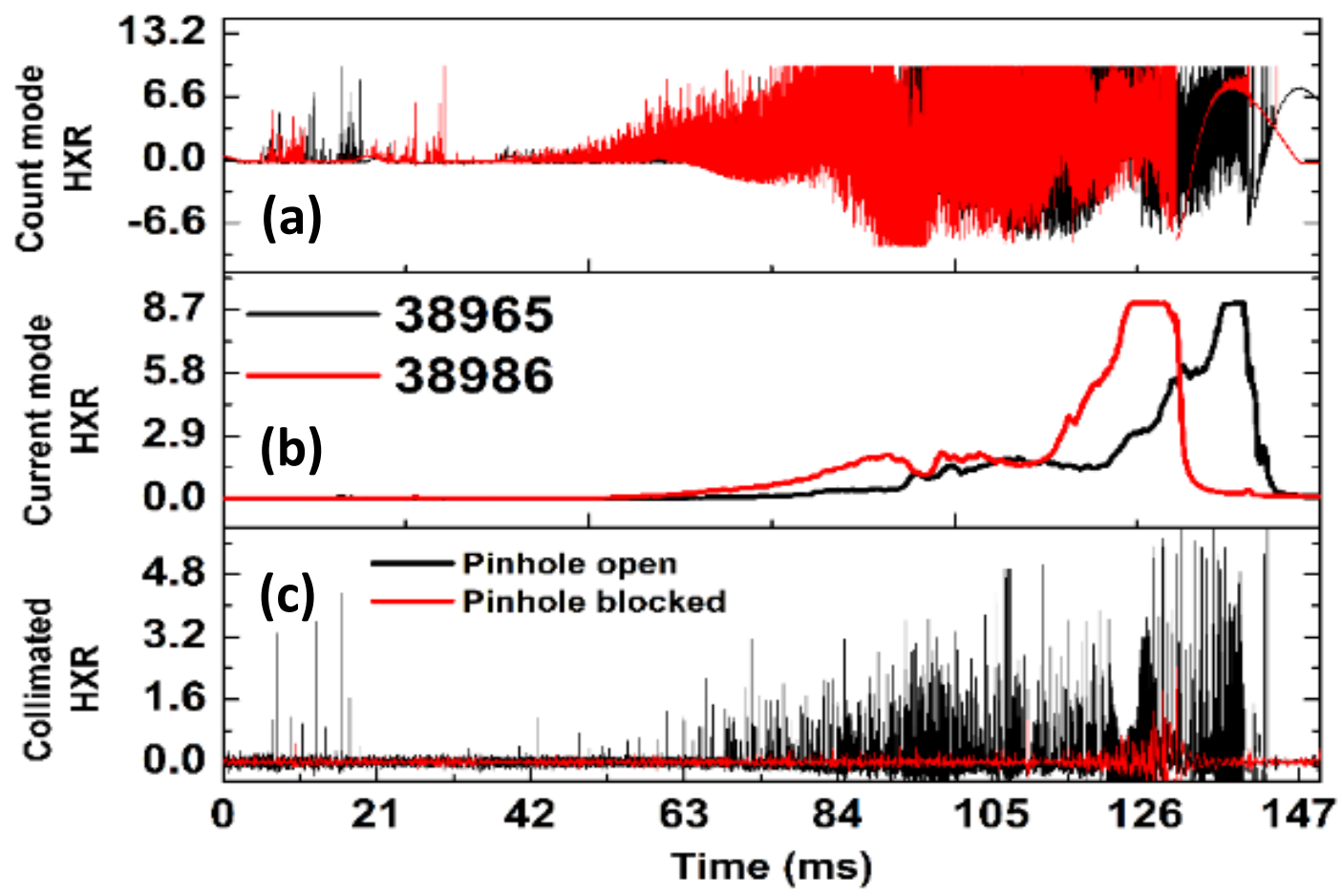

**Figure 4.** Hard X-ray (HXR) emissivity measured using different diagnostics: (a) and (b) combined thick- and thin-target bremsstrahlung emission recorded by a NaI scintillation detector operating in count mode and current mode, respectively; (c) thin-target bremsstrahlung emission measured using a CdTe detector in count mode.

Furthermore the experimental results have been validated with a forward modeling code reported in references[15]. The code has been utilized to evaluate the total photon flux at the detector. In the thin-target approximation, the local bremsstrahlung emissivity is governed by several key parameters, including the runaway electron (RE) density $n_{RE}$, the background plasma (or target ion) density $n_e$, the RE momentum distribution function $f(p)$, and the atomic number Z of the interacting species. A central quantity in this formulation is the double differential cross-section (DDCS, $\frac{\mathrm{d}^2\sigma_{ff}}{\mathrm{d}E_\gamma \mathrm{d}\Omega_\gamma}$, which characterizes the probability per unit photon energy interval $dE_\gamma$ and per unit solid angle $d\Omega_\gamma$ for photon emission resulting from free–free interactions between electrons and plasma species such as ions and electrons. The DDCS depends explicitly on the RE energy ($E_{RE}$), the photon emission angle ($\theta_0$), and atomic charge (Z) of interaction ion. Here, $\theta_0$ denotes the angle between the electron momentum vector and the line of sight to the detector. This angle is intrinsically related to the pitch angle $\theta_p$, defined as the angle between the RE momentum and the toroidal magnetic field $B_T$. Consequently, the bremsstrahlung flux detected is highly sensitive to both the electron energy distribution and the pitch-angle distribution of the runaway electrons[22].

$$\frac{dN(E_\gamma)}{dE_\gamma dt} = \eta_A(E_\gamma)\eta_f(E_\gamma)\eta_p(E_\gamma)A_{det}\Omega_{det} \times \int dl n_i n_{RE} \int f(p) d^3p\, \beta c \times \left[\sum_i \left(\frac{n_i}{n_e}\right)\frac{d^2\sigma_{ei}}{dE_\gamma d\Omega_\gamma} + \frac{d^2\sigma_{ee}}{dE_\gamma d\Omega_\gamma} + \frac{d^2\sigma_{DRR}}{dE_\gamma d\Omega_\gamma}\right] \quad (1)$$

The quantity $\frac{dN(E_\gamma)}{dE_\gamma dt}$ denotes the number of bremsstrahlung $\gamma$-photons per unit photon energy interval, $E_\gamma$ to $E_\gamma + dE_\gamma$, per unit time, dt, entering into the detector volume, as defined in Eq.1. In this formulation, $\eta_p(E_\gamma)$ represents the transmission efficiency of photons through the plasma. For optically thin plasmas, typically valid for $E_\gamma \gg 0.01\ MeV$, this factor can be approximated as unity. Since the detector is positioned outside the vacuum vessel and views the plasma through a glass viewport, photon attenuation due to the viewport material must be considered. Additionally, photons traversing from the plasma to the detector encounter various intervening materials, including shielding components and the detector window. The cumulative transmission efficiency of these materials is described by $\eta_f(E_\gamma)$.The intrinsic detection efficiency of the detector is denoted by $\eta_A(E_\gamma)$ which depends sensitively on photon energy and detector thickness. The effective detector area exposed to the incoming photon flux is given by $A_{det}$, while $\Omega_{det}$ defines the detector solid angle, determined by the geometry of the collimation system, specifically its aperture and length. The first integral term in Eq. (1) is evaluated along the entire line-of-sight (LoS) chord length l of the detector. The terms $\frac{d^2\sigma_{ei}}{dE_\gamma d\Omega_\gamma}$ and $\frac{d^2\sigma_{ee}}{dE_\gamma d\Omega_\gamma}$ correspond to the double differential cross-sections (DDCS) for electron-ion (e-i) and electron-electron (e-e) bremsstrahlung processes, respectively, with units of $m^2 MeV^{-1} sr^{-1}$. The final term in the equation accounts for the DDCS associated with direct radiative recombination (DRR). For photon energies exceeding $0.1 MeV$ and relativistic electron energies ($> 1\ MeV$), the contribution of DRR is negligible compared to e-i bremsstrahlung and is therefore often omitted in practical calculations. Finally, the number of bremsstrahlung photons recorded by the pulse-height analysis system within the detected energy interval, $E'_\gamma$ to $E'_\gamma + dE'_\gamma$, and per unit time, dt, can be determined using the corresponding detector response relation, as expressed in Eq.2.

$$\frac{\mathrm{d}N(E'_\gamma)}{\mathrm{d}E'_\gamma \mathrm{d}t} = \int_{E_\gamma} \mathrm{d}E_\gamma D_{\mathrm{det}}(E_\gamma, E'_\gamma) \frac{\mathrm{d}N(E_\gamma)}{\mathrm{d}E_\gamma \mathrm{d}t} \tag{2}$$

The detector response function, $D_{\mathrm{det}}(E_\gamma, E'_\gamma)$, depends on several factors, including the incident photon energy, as well as the detector material, geometry, and dimensions. Photon interactions within the detector volume occur through multiple mechanisms, primarily the photoelectric effect, Compton scattering, and pair production. In the low-energy regime, the photoelectric effect is the dominant interaction process, resulting in nearly complete deposition of the photon energy within the detector. For a CdTe semiconductor detector with an active area of $25\ mm^2$ and a thickness of 2 mm, the effective detection range typically spans photon energies of approximately $20 - 350\ keV$. Within this energy interval, the response function can be approximated such that, $\int_{E_\gamma} \mathrm{d}E_\gamma D_{\mathrm{det}}(E_\gamma, E'_\gamma) = 1$, reflecting the predominance of full-energy absorption via the photoelectric effect. To further simplify the analysis, a monoenergetic runaway electron (RE) beam with negligible momentum distribution is assumed in Eq.1. Under these conditions, and for a pure hydrogen $H_2$ plasma without impurity species, the bremsstrahlung photon spectrum measured by the pulse-height detection system can be expressed in the simplified form given by Eq.3.

$$\frac{\mathrm{d}N(E'_\gamma)}{\mathrm{d}E'_\gamma \mathrm{d}t} = \eta_A(E_\gamma)\eta_f(E_\gamma)A_{det}\Omega_{det} \times L \times n_i \times n_{RE}\left[\frac{d^2\sigma_{ei}}{dE_\gamma d\Omega_\gamma} + \frac{d^2\sigma_{ee}}{dE_\gamma d\Omega_\gamma}\right] \tag{3}$$

Here, L denotes the total chord length along the detector’s line of sight (LoS). The plasma density and runaway electron (RE) density are assumed to be spatially uniform across the entire plasma volume. The principal design parameters of the thin-target bremsstrahlung diagnostic system are summarized in Table 1.

**Table 1.** Key design parameters of the thin-target bremsstrahlung detection system, along with the corresponding plasma characteristics under thin-target conditions.

| Parameter | Value |
|---|---|
| **Detector type** | **CdTe** |
| **Detector thickness** | $\mathbf{2\ mm}$ |
| **Detector active area** | $\mathbf{25\ mm^2}$ |
| **Energy resolutions** | $\mathbf{\sim 6 - 8\ keV\ (at\ 122\ keV)}$ |
| **Typical energy range** | $\mathbf{\sim 20 - 350\ keV}$ |
| **Detector window** | **Al, $\mathbf{1.5\ mm}$** |
| **Collimator length** | **Lead, $\mathbf{320\ mm}$** |
| **Collimator diameter and aperture** | **Diameter- $\mathbf{820\ mm}$; Aperture- $\mathbf{8\ mm}$** |
| **LOS length (L)** | $\mathbf{0.50}$ **m** |

| Detector's FOV ($\Omega$) | 0.002 sr |
|---|---|
| Plasma current ($I_p$) | 150 kA |
| View port | Fused silica, 8 mm thick |
| Runaway current ($I_{RE}$) | (1-2)% of $I_p$ ~2 kA |
| Spatially uniform RE density ($n_{RE}$) | $2 \times 10^{14} m^{-3}$ |
| Spatially uniform plasma density ($n_i$) | $1 \times 10^{18} m^{-3}$ |

In the Aditya-U tokamak, runaway electrons (REs) are predominantly generated during the plasma start-up phase. The inductive start-up is characterized by a relatively high loop voltage and low plasma density, conditions that favor the generation of REs primarily through the Dreicer mechanism. In contrast, the exponential multiplication of REs via the avalanche process is expected to be weak under these conditions. This is due to the reduced loop voltage following plasma breakdown, as well as the comparatively long avalanche growth timescale relative to the overall plasma duration. Consequently, the contribution of runaway electrons to the total plasma current is assumed to remain limited. Based on observations reported in similar tokamak experiments[25,26], the runaway current fraction is estimated to be on the order of 1–2% of the total plasma current.

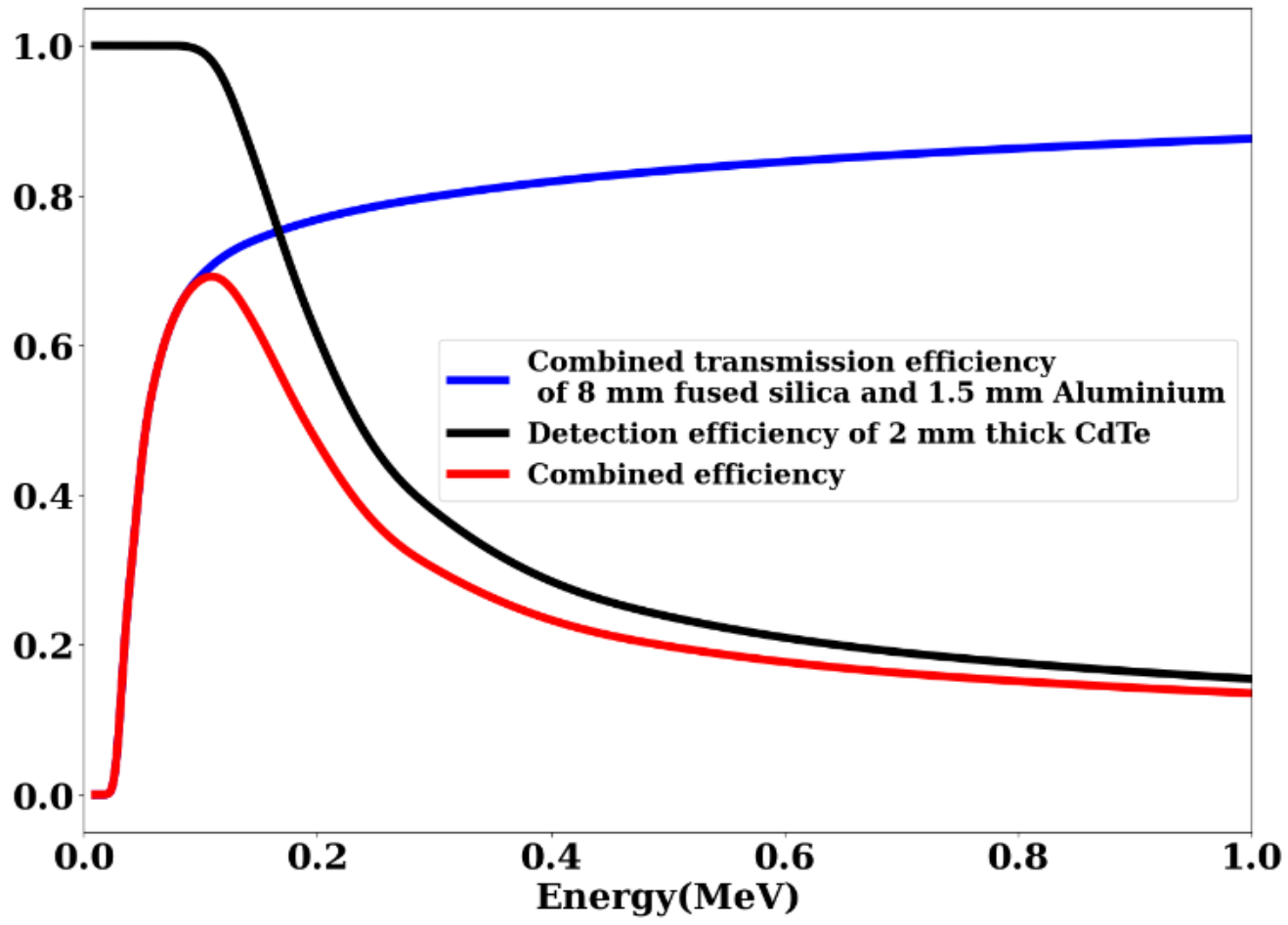

Figure 5. Detection efficiency of 2 mm thick CdTe detector is represented by black trace, blue trace is the combined transmission efficiency of 8 mm thick fused silica view port and 1.5 mm thick Al window, red trace represents the fraction of $\gamma$-photons those interact within the detector volume and emitted from the plasma.

Figure 5 illustrates the transmission efficiencies of the various materials encountered along the path between the plasma emission region and the detector. The detection efficiency of a CdTe detector is strongly dependent on its thickness[27]. In the present experiment, a CdTe detector with a thickness of 2 $mm$ is employed, and its corresponding efficiency is shown in Fig. 5. As the detector thickness increases, higher-energy $\gamma$ photons exhibit a greater probability of interaction within the detector volume. In contrast, the interaction probability for low-energy photons tends to saturate beyond a certain thickness. Using the model equation (Eq. 3) and the parameter values listed in Table 1, the bremsstrahlung photon counts recorded by the pulse-height analysis system over a time interval of 60 ms have been evaluated. As discussed previously, the double differential cross-section (DDCS) depends on the atomic number Z, the runaway electron energy $E_{RE}$, and the pitch angle $\theta_p$. For typical Aditya-U plasma conditions, the effective charge is estimated to be Z~2, as inferred from visible continuum measurements[28]. The dependence of the DDCS on beam energy and pitch angle is presented in Fig. 6. For $\theta_p = 0^0$ the corresponding photon emission angle relative to the detector line of sight is $\theta_0 = 90^0$. Figure 6a indicates that, for a fixed pitch angle, the photons reaching the detector decreases with increasing beam energy. Similarly, for a fixed beam energy of 1 $MeV$, the detected photons number decreases as the pitch angle increases (see Fig.6b).

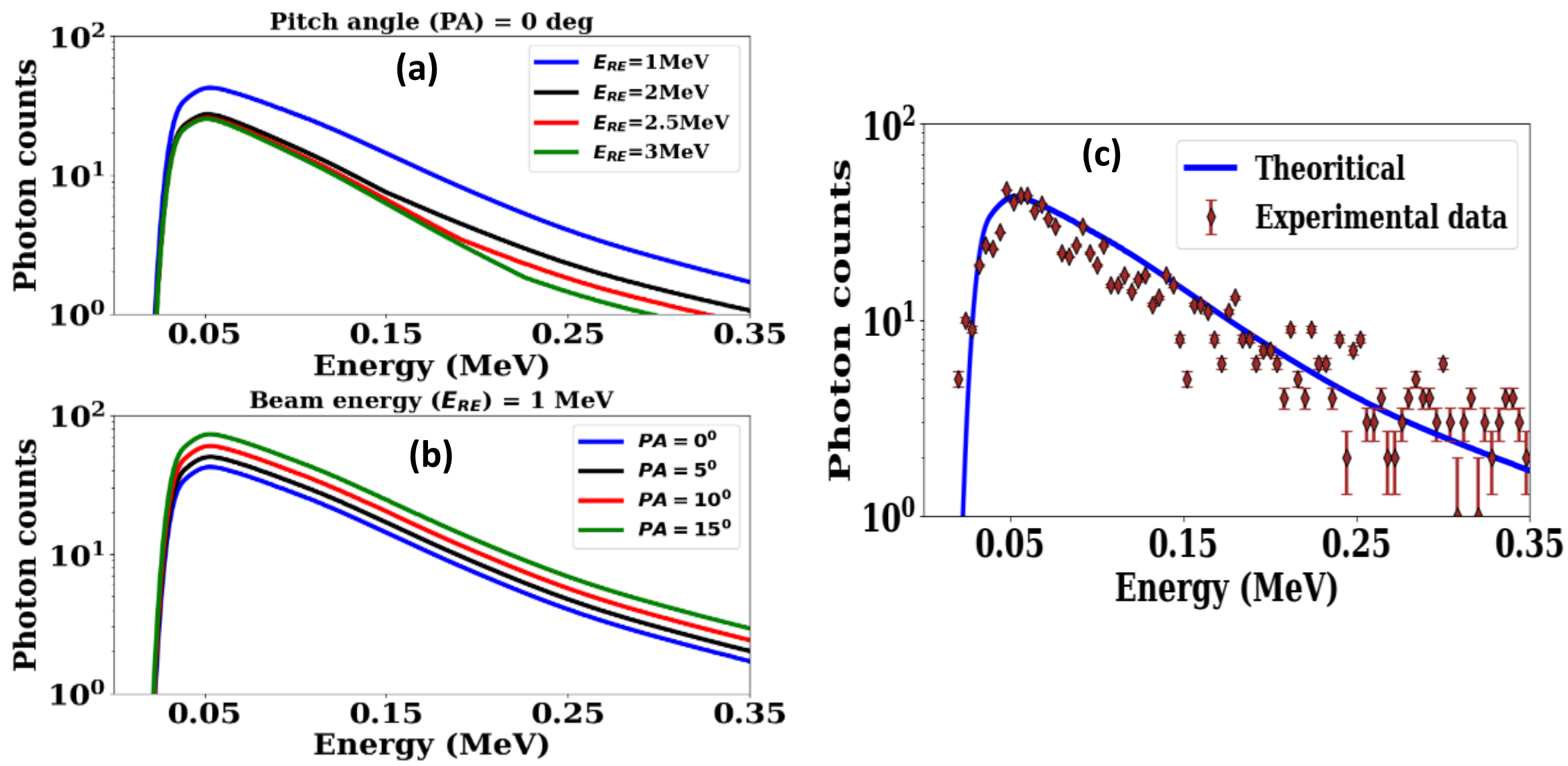

Figure 6. Forward modelling illustrating the sensitivity of the HXR photon energy distribution to (a) runaway electron (RE) beam energy and (b) pitch angle. The experimentally obtained pulse-height distribution of HXR photons is compared with synthetic spectra from the forward-modelling code, assuming an RE beam energy of 1 $MeV$ and a pitch angle of 0°.

The hard X-ray (HXR) signal acquired through the collimator with the aperture in the open configuration is shown in Fig. 4c as the black trace. Pulse-height analysis of this count-mode signal was performed over a 60 ms time interval ($t = 63 - 123\, ms$) by 4 keV energy binning the recorded data. As presented in Fig.6c, the experimentally measured HXR spectrum exhibits good agreement with the theoretically predicted spectrum when a runaway electron (RE) beam energy of 1 MeV and a pitch angle of $0^0$ are assumed. The experimental data points are accompanied by vertical error bars, calculated as $1/\sqrt{N}$, where N is the photon count; this corresponds to the statistical uncertainty arising from Poisson-distributed counting noise.

## IV. Conclusion:

In the Aditya-U tokamak, thin-target hard X-ray (HXR) bremsstrahlung emission has been measured for the first time using a specially designed diagnostic system consisting of a shielded CdTe detector integrated with a lead collimator. This diagnostic has been developed to study confined runaway electrons. CdTe is advantageous over PMT-based detectors for this application, as it allows the setup to be placed very close to the plasma, which is essential for obtaining a field-of-view-restricted signal. Although alternative configurations using scintillation detectors with remotely coupled PMTs via optical components can also allow detector placement near the plasma, such arrangements are more complex[29]. In contrast, the CdTe-based system provides a compact and efficient solution with high sensitivity in the photon energy range of 20–350 keV. This makes it particularly suitable for investigating non-thermal electron populations, such as those generated during sawtooth activity, typically in the energy range of 200–350 keV. The combined use of the newly developed thin-target bremsstrahlung diagnostic and the existing HXR monitoring system, which detects both thin-target and limiter-target emission, provides a powerful tool to address longstanding questions regarding the origin of HXR spikes during sawtooth events. In particular, the observation of HXR spikes in the limiter-viewing system, correlated with sawtooth crashes, alongside their absence in the thin-target diagnostic, would indicate that sawtooth crashes predominantly lead to the loss of pre-existing REs rather than the generation of new non-thermal electrons. In addition the hxr spectra generated from thin target hxr signal can estimate the energy range and pitch angle of the confined RE beam. In future a suitable distribution function of REs will be chosen judiciously and will incorporated in the forward modelling code. Thus, the impact of both the runaway electron (RE) distribution and the spatial distribution of plasma density on the resulting hard X-ray (HXR) spectra will be examined. Hence with these additional considerations, the best match between theoretically and experimentally obtained hxr spectra will enable us to estimate precise RE beam energy and pitch angle.